\documentclass[usenatbib,usegraphicx]{mn2e}
\usepackage{subfigure}
\usepackage{longtable}
\usepackage{graphicx}
\usepackage{graphics}
\usepackage{keyval}
\usepackage{trig}
\usepackage{psfig,epsf}
\usepackage{times}
\def\beq{\begin{equation}}
\def\eeq{\end{equation}}
\newcommand{\ltsima}{$\; \buildrel < \over \sim \;$}
\newcommand{\lesssim}{\lower.5ex\hbox{\ltsima}}
\newcommand{\gtrsim}{\lower.7ex\hbox{$\;\stackrel{\textstyle>}{\sim}\; 
$}} 

\title{Mass models from high-resolution HI data of the dwarf galaxy NGC 1560}
\author[G. Gentile, M. Baes, B. Famaey, K. Van Acoleyen]
{G. Gentile$^{1}$, M. Baes$^{1}$, B. Famaey$^{2}$, K. Van Acoleyen$^{1}$ \\
$^{1}$Department of Physics and Astronomy, Universiteit Gent, Krijgslaan 281, B-9000
  Gent, Belgium\\
$^{2}$Observatoire Astronomique, Universit\'e de Strasbourg, CNRS UMR 7550,
  F-67000 Strasbourg, France
}

\begin{document}

\date{Accepted ... Received ... ; in original form ...}

\pagerange{\pageref{firstpage}--\pageref{lastpage}} \pubyear{2010}

\maketitle

\label{firstpage}

\begin{abstract}

We present HI observations performed at the GMRT of the 
nearby dwarf galaxy NGC 1560. This Sd galaxy is well-known
for a distinct ``wiggle'' in its rotation curve. Our 
new observations have twice the resolution of the previously
published HI data. We derived the rotation curve by taking
projection effects into account, and we verified the derived
kinematics by creating model datacubes. This new rotation 
curve is similar to the previously published one: we confirm
the presence of a clear wiggle. The main differences are
in the innermost $\sim 100 \arcsec$ of the rotation curve, 
where we find slightly ($\lesssim 5$ km s$^{-1}$) higher velocities.   
Mass modelling of the rotation curve results in good fits using
the core-dominated Burkert halo (which however does not reproduce
the wiggle), bad fits using the a Navarro, Frenk \& White halo, 
and good fits using MOND (Modified
Newtonian Dynamics), which also reproduces the wiggle.

\end{abstract}


\section{Introduction}

Rotation curves of spiral galaxies are one of the most important tools
to investigate the content and distribution of dark matter in galaxies.
They have been used for a variety of purposes, in particular to 
investigate their systematic properties (Persic, Salucci \& Stel 1996, 
Salucci et al. 2007, Gentile 2008), to test the validity of the 
predictions of the Cold Dark Matter (CDM) theory (e.g., de Blok et al. 2001,
Marchesini et al. 2002, Gentile et al. 2004, 2005, 2007a, Kuzio de Naray et
al. 2006, Corbelli et al. 2010), or to study the connection 
between the distributions of dark
and luminous matter (Broeils 1992, McGaugh et al. 2000, McGaugh 2005a, 
Gentile et al. 2009, Donato et al. 2009).

Testing the validity of the CDM predictions is a very important 
issue, because (CDM-only) simulations result in dark matter halos
with an almost universal density profile (the details of how universal
the profile actually is have extensively been discussed in the literature),
which is well described by the NFW (Navarro, Frenk \& White 1996) halo,
characterised by a central density cusp (the density $\rho$ is
proportional to $r^{-1}$ when the radius $r$ tends to zero, but see
also Section \ref{sect_results}),
whereas observations tend to indicate the presence of a constant
density core.
The influence of baryons on the distribution of dark matter is a crucial
point; however, there is no general consensus about what the dominant effect(s)
is(are). A non-exhaustive list of ways by which baryons can change 
the distribution of dark matter includes: adiabatic contraction (Blumenthal
et al. 1986, Gnedin et al. 2004, Sellwood \& McGaugh 2005), the influence of 
bars (e.g., Weinberg \& Katz 2002, McMillan \& Dehnen 2005, Sellwood 2008), or the influence of gas 
(Mashchenko, Couchman \& Wadsley 2006, Governato et al. 2010).
Because of the additional complication brought by baryons, dwarf
galaxies and low surface brightness (LSB) galaxies are better suited
for deriving the properties of dark matter in galaxies.

An alternative explanation to dark matter in galaxies is MOND
(Modified Newtonian Dynamics, introduced by Milgrom 1983), where
the effective gravitational acceleration becomes stronger than expected
in the Newtonian case, when the gravitational acceleration falls
below a critical value, $a_0 \sim 1.2 \times 10^{-8}$ cm s$^{-2}$.
MOND explains very well the observed kinematical properties
of galaxies: LSB galaxies (McGaugh \& de Blok 1998),
tidal dwarf galaxies (Gentile et al. 2007b), the Milky Way 
(Famaey, Bruneton \& Zhao 2007a, McGaugh 2008, Bienaym\'e et al. 2009), 
early-type spiral galaxies (Sanders \& Noordermeer 2007),
elliptical galaxies (Milgrom \& Sanders 2003, Tiret et al. 2007), 
and galaxy scaling relation
in general, such as the baryonic Tully-Fisher relation (McGaugh 2005b).
However, let us note that the MOND prescription is not sufficient to 
explain the observed discrepancy between luminous and dynamical mass 
in galaxy clusters (e.g. Angus et al. 2007).

The dwarf galaxy NGC 1560, whose rotation curve was derived
by Broeils (1992, hereafter B92) based on WSRT (Westerbork Synthesis Radio
Telescope) observations, is a nearby Sd galaxy with an absolute B-band
magnitude of $M_{\rm B}=-16.6$ (Krismer et al. 1995, assuming a distance
of 3.5 Mpc). Estimates of
its distance vary significantly from one study to another: values 
from 2.5 Mpc (Lee \& Madore 1993) to 3.7 Mpc (Sandage 1988) can be 
found. B92 assumed a distance of 3 Mpc. Karachentsev et al.
(2003) find $3.45 \pm 0.36$ Mpc from the tip of the red giant branch method, 
using HST data. We assume this value unless stated otherwise, because it
is one of the most accurate to date. 

NGC 1560 is a very interesting
galaxy to study because it is the stereotypical galaxy displaying
what is known as ``Renzo's rule'' (from Sancisi 2004): for every
feature in the distribution of visible matter there is a corresponding
feature in the total distribution of matter. In the rotation curve
of NGC 1560, as derived by B92, there is a clear ``wiggle'' in the total
rotation velocity, which corresponds very closely to a similar 
wiggle in the gas contribution to the rotation curve. 
Mass models such as MOND naturally reproduce the feature, whereas
models that include a dominant spherical (or triaxial) halo
are too smooth to do so. This motivated
us to reobserve NGC 1560 at higher resolution, to independently
trace the rotation curve and probe the region around the velocity
wiggle.

In the present paper, we present an analysis of HI observations
performed with the GMRT (Giant Metrewave Radio Telescope) which
have a spatial resolution almost two times better than the data
presented in B92. We re-derive the rotation curve and make
mass models, with various assumptions concerning the 
distribution of dark matter.

\begin{figure*}
\begin{center}
\includegraphics[scale=1.15]{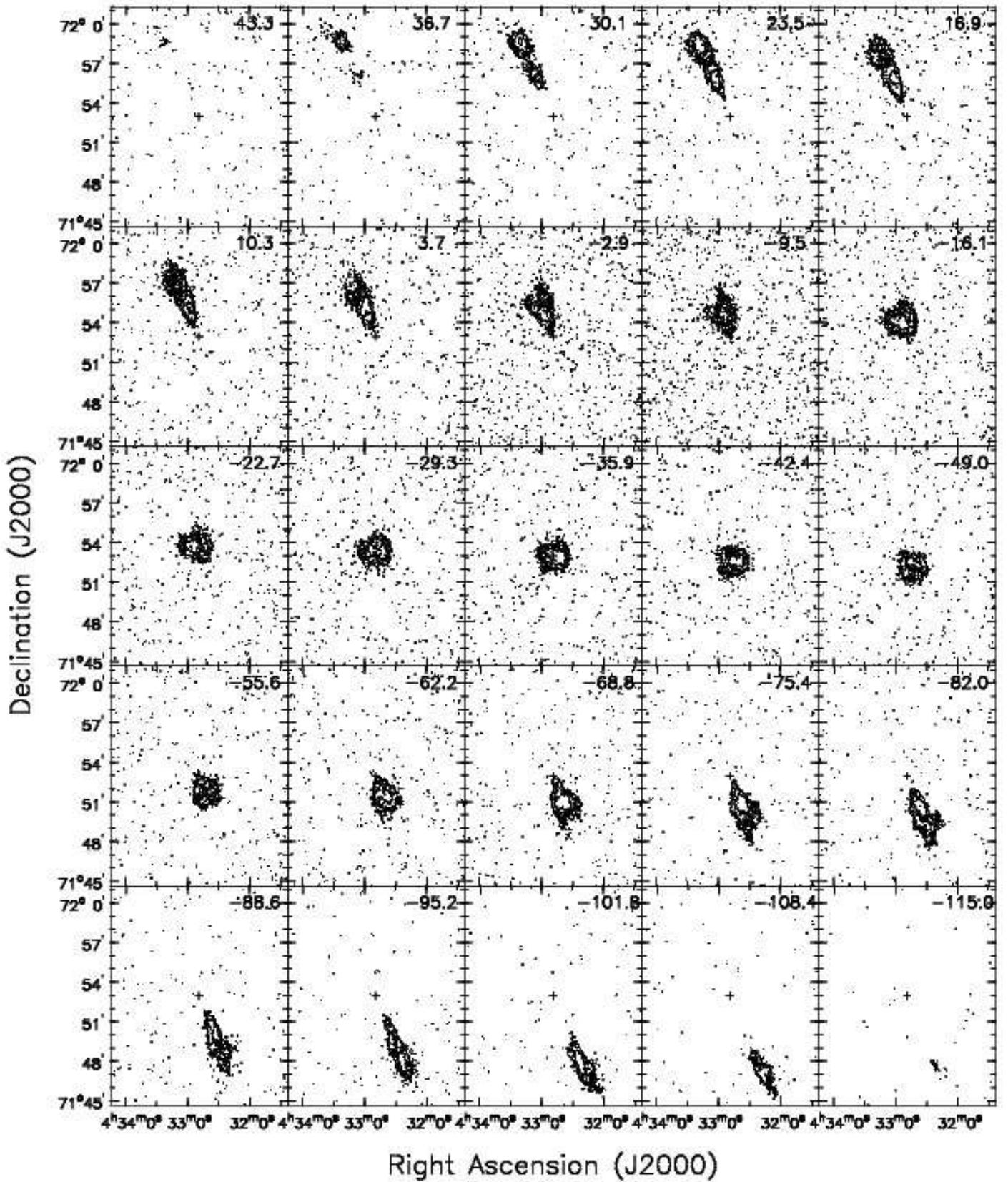}
\end{center}
\caption{Observed channel maps of NGC 1560. The heliocentric radial
velocity (in km s$^{-1}$) is indicated at the top right corner of
each channel map. Contours are -2.5, 2.5 ($\sim 3 \sigma$), 
5, 10, 20, 40 mJy
beam$^{-1}$. The cross shows the location of the galaxy centre. 
The synthesized beam is 8.1$ \arcsec \times 6.4 \arcsec $.
} \label{channels_hires}
\end{figure*}

\section{observations}

\begin{table}
\caption{\label{obs-table} Observational parameters of the
HI observations with the GMRT.}
\begin{tabular}{ll}
\hline
Observing dates     & 13-14 Sep 2007             \\
Time on source (mins)     & 690            \\
Synthesized beam & 8.1$ \arcsec \times 6.4 \arcsec $\\
Number of velocity channels & 128\\
Velocity increment &    6.7 km s$^{-1}$     \\
rms noise in the channel maps &    0.8 mJy beam$^{-1}$     \\
\hline
\end{tabular}
\end{table}

The observations were performed on 13-14 September 2007 at 
the GMRT. The correlator setup was such that the total
bandwidth was 2 MHz, with 128 channels centred around 
a heliocentric (optical definition) systemic velocity
of -36 km s$^{-1}$. After Hanning smoothing, the velocity
resolution of our datacube is 6.7 km  s$^{-1}$. 
Standard calibration and editing procedures were performed
within the AIPS (Astronomical Image Processing System) 
software package.
The absolute flux and bandpass 
calibration were performed using the standard calibrators
3C 48 and 3C 147, whereas the phase calibrator was chosen
to be 0410+769. After calibration, the data were 
continuum-subtracted using line-free channels on either
side of the channels with line emission.

Imaging was performed using the task IMAGR in AIPS. To
avoid resolving excessively the extended structure, after
various attempts we decided that a Gaussian taper of 25 k$\lambda$
provides a good compromise between resolution and sensitivity to
extended emission. The synthesized beam of our final high-resolution
maps is 8.1$ \arcsec \times 6.4 \arcsec $, which is almost a factor
of 2 better than B92 (whose beam size was 14$ \arcsec \times 13 \arcsec$). 
In the first maps we produced, we noticed the
presence of a weak ``negative bowl'' around the emission, characteristic
of missing short-spacing information. However, following Greisen,
Spekkens \& van Moorsel (2009), we used the technique of multi-scale
CLEAN, which they found can eliminate most of the negative flux
around the emission, and thus solve almost completely the 
missing short-spacing problem. As we explain in the next section,
the use of multi-scale CLEAN likely enabled us to recover 
most of the HI flux of NGC 1560.

\begin{figure*}
\includegraphics[scale=0.7]{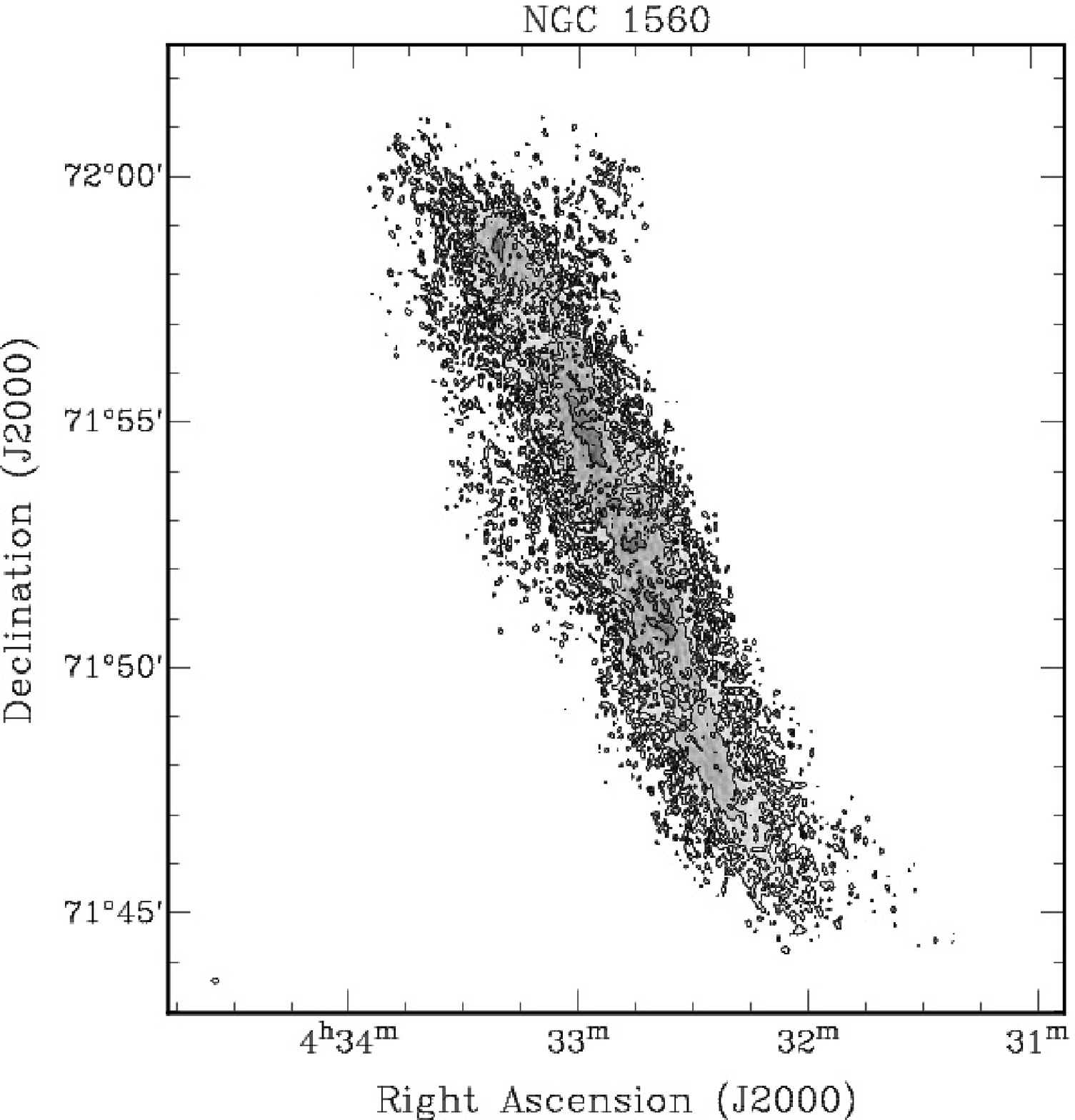}
\caption{Total HI map based on the high-resolution datacube. 
Contours are (8, 16, 32, 64)$\times 10^{20}$ atom cm$^{-2}$.
The lowest contour represents the ``pseudo-3$\sigma$'' level
defined in the same way as Verheijen \& Sancisi (2001).
The synthesized beam is 8.1$\arcsec \times 6.4 \arcsec$.
}
\label{mom0_hires}
\end{figure*}

\begin{figure*}
\includegraphics[scale=0.4]{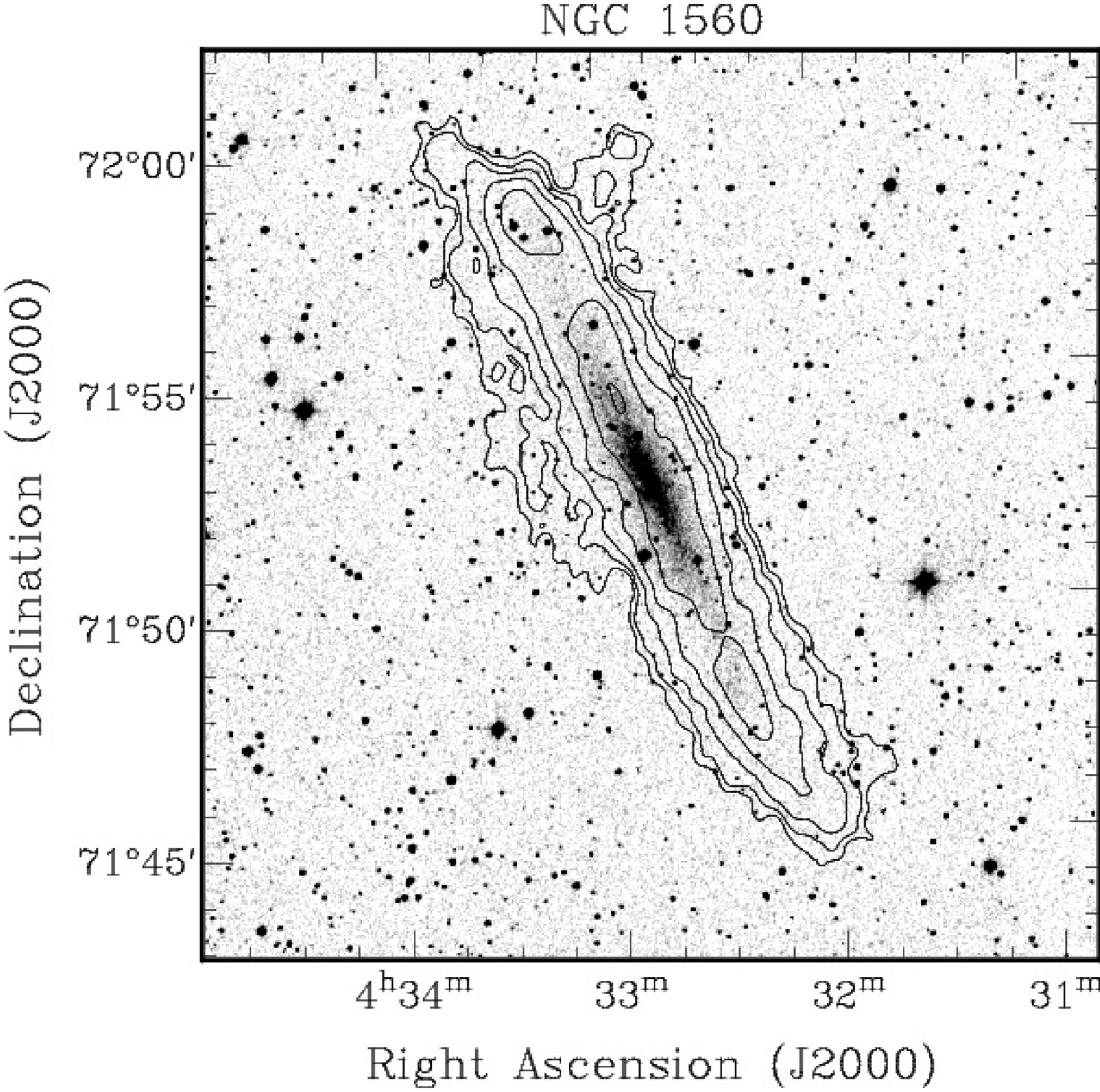}
\caption{Total HI map based on the low-resolution datacube,
overlaid with an optical (DSS) image.  
Contours are (1, 2, 4, 8, 16, 32, 64)$\times 10^{20}$ atom cm$^{-2}$).
The lowest contour represents the ``pseudo-3$\sigma$'' level
defined in the same way as Verheijen \& Sancisi (2001).
The synthesized beam is 25$\arcsec \times 25 \arcsec$.
}
\label{mom0_lores_opt}
\end{figure*}

\begin{figure*}
\includegraphics[scale=0.6]{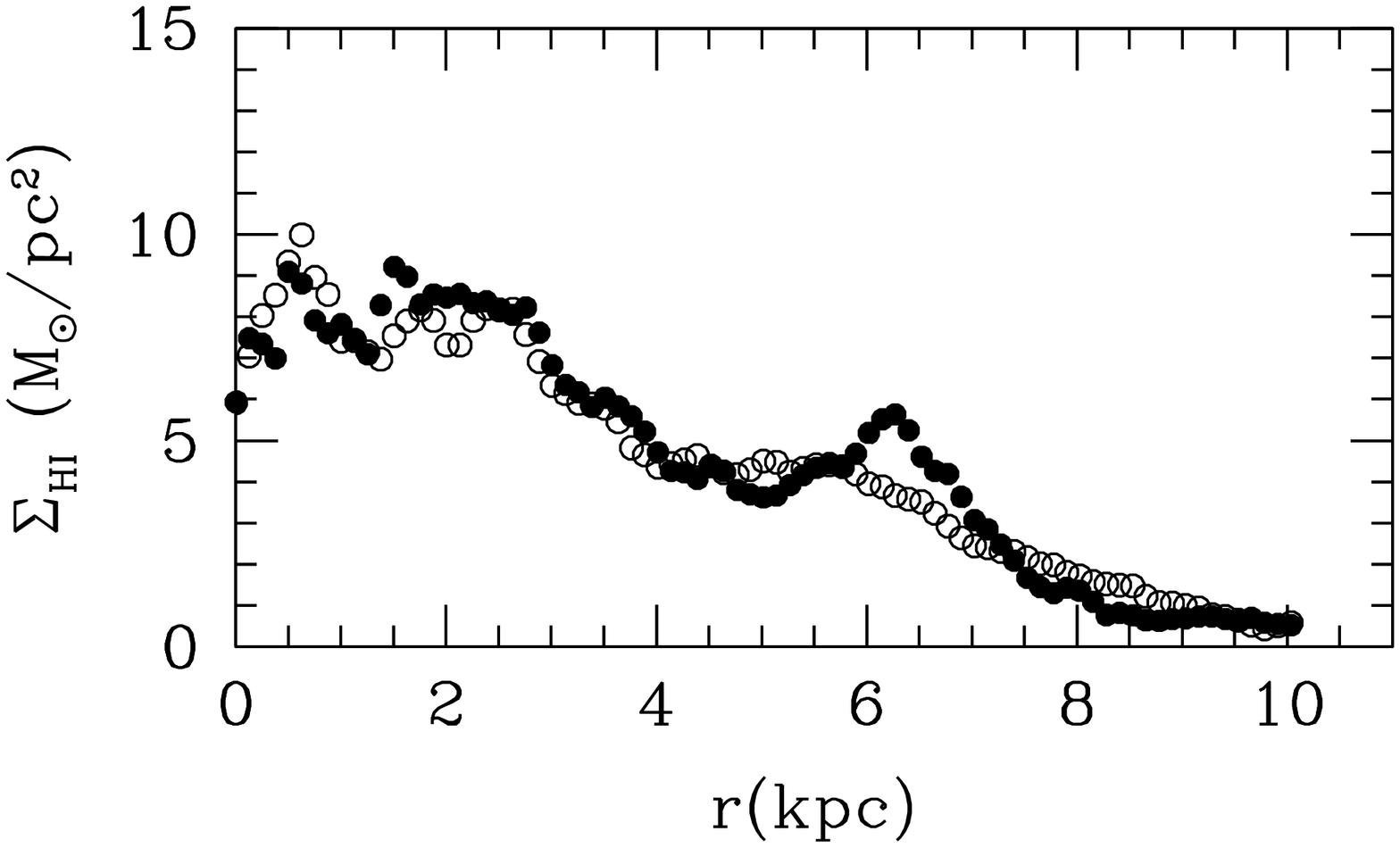}
\caption{HI surface density, calculated from averaging over ellipses,
as a function of radius. Full/empty circles represent the northern/southern 
half of the galaxy.
} \label{surfdens}
\end{figure*}

\section{HI in NGC 1560}
\label{sect_hi}

The final high-resolution datacube is shown in Fig. \ref{channels_hires}.
One can notice that the emission traces the rotation of a highly inclined
(but not fully edge-on) disk. 
The rms noise in the channel maps is 0.8 mJy beam$^{-1}$.
A few channels around zero velocity
have significantly higher noise, which can be explained by the presence of 
very diffuse, low-surface brightness emission due to galactic HI.

The total HI flux (calculated from the primary beam corrected low-resolution
cube) is 294.6 Jy km s$^{-1}$, which is 23 \% lower than 
B92, but is consistent with the single-dish total flux of 290 Jy km s$^{-1}$
given by de Vaucouleurs et al. (1991).  
Importantly, the total flux of our HI datacube
derived with the point-source CLEAN would have been 32 \% lower, showing the 
ability of multi-scale CLEAN to deal with short spacing data. 
The total HI map (moment-0 map) is given in Fig. \ref{mom0_hires}.

To better study the extended, low surface brightness emission, we constructed
a low-resolution datacube with a beam of $25 \arcsec \times 25 \arcsec$. The
resulting total HI map, superimposed with an optical image of NGC 1560,
is shown in Fig. \ref{mom0_lores_opt}.

We derived the surface density profile by averaging over ellipses
using the geometrical parameters derived in the next section. Small
changes in the geometrical parameters do not affect significantly 
the resulting surface density profile, which is shown in Fig. 
\ref{surfdens}. Similarly to what B92 found, the HI distribution 
is quite symmetric, apart from the ``bump'' around 300-350 $\arcsec$,
which is very promiment on the northern side and just hinted at on
the southern side.

\section{Rotation curve}

\begin{figure*}
\includegraphics[scale=0.7]{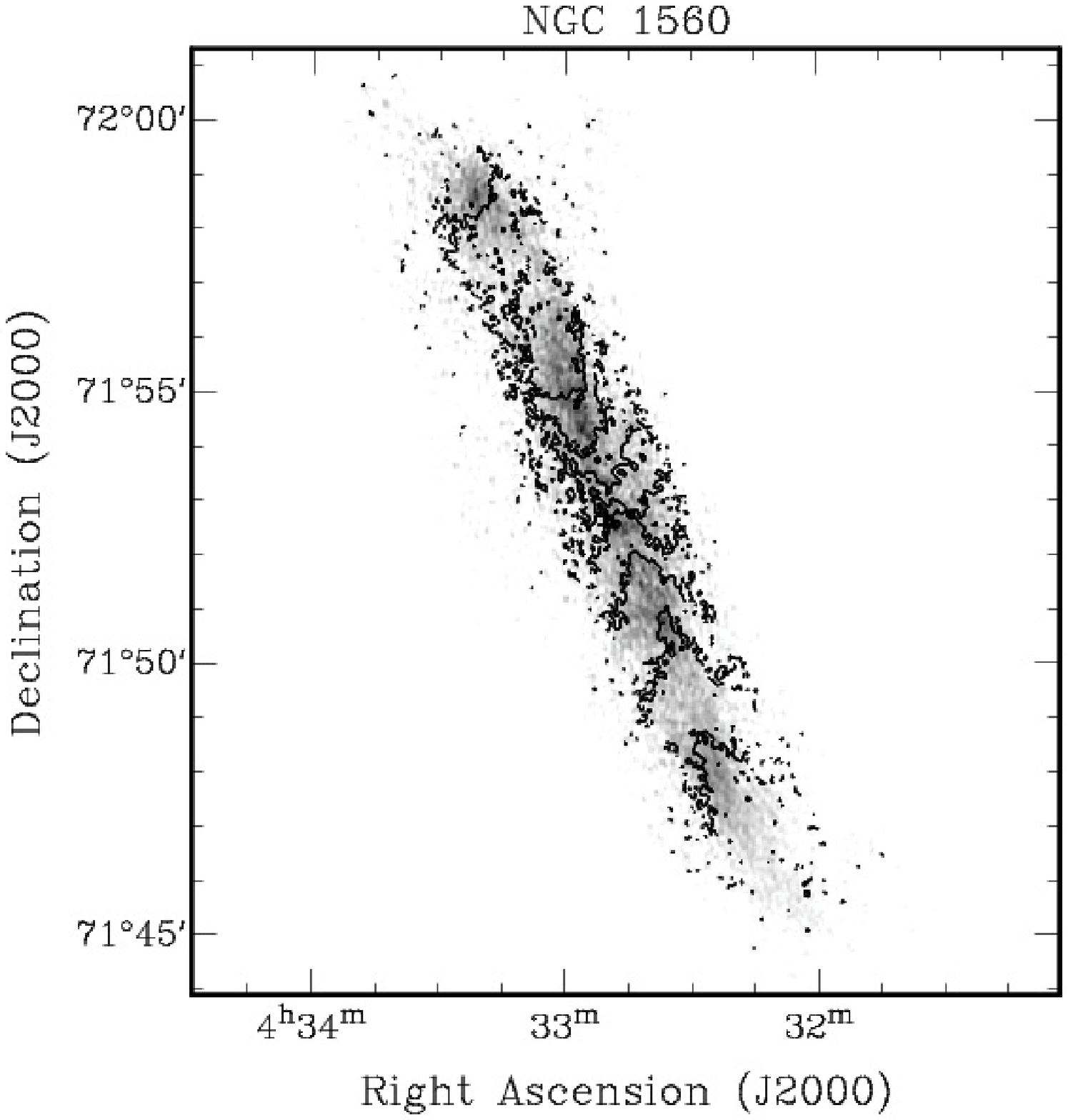}
\caption{High-resolution total HI map (greyscale) and velocity 
field (contours). Contours are centred around $-36$ km s$^{-1}$ and 
spaced by 15 km s$^{-1}$.}
\label{mom0_hires_velfi}
\end{figure*}

The velocity field of NGC 1560 was derived using the WAMET method
(Gentile et al. 2004, where the velocity in each position of the velocity
field is derived using only the side of the velocity profile opposite to the 
systemic velocity), which gives better results than traditional
methods (such as the intensity-weighted mean) when projection and/or resolution
effects are expected to be non-negligible. In the case of NGC 1560, because
of its high inclination angle, projection effects could potentially bias towards
lower rotation velocities (see e.g. Sancisi \& Allen 1979), therefore we 
decided to use the method described in Gentile et al. (2004) instead of
the intensity-weighted mean. The velocity
field, overlaid with the total HI map, is shown in Fig. \ref{mom0_hires_velfi}.

Once the velocity field was constructed, we derived the rotation curve
using the task ROTCUR in GIPSY (Groningen Imaging Processing System), which 
makes a tilted-ring model of
the velocity field (Begeman 1989). Several attempts were made to leave as many free
parameters as possible, and at the same time have stable solutions for
the rings with enough points. We ended up leaving as free parameters
(apart from the rotation velocity) the position angle and the systemic
velocity. The inclination was fixed at its average value. 

\begin{figure*}
\includegraphics[scale=0.7]{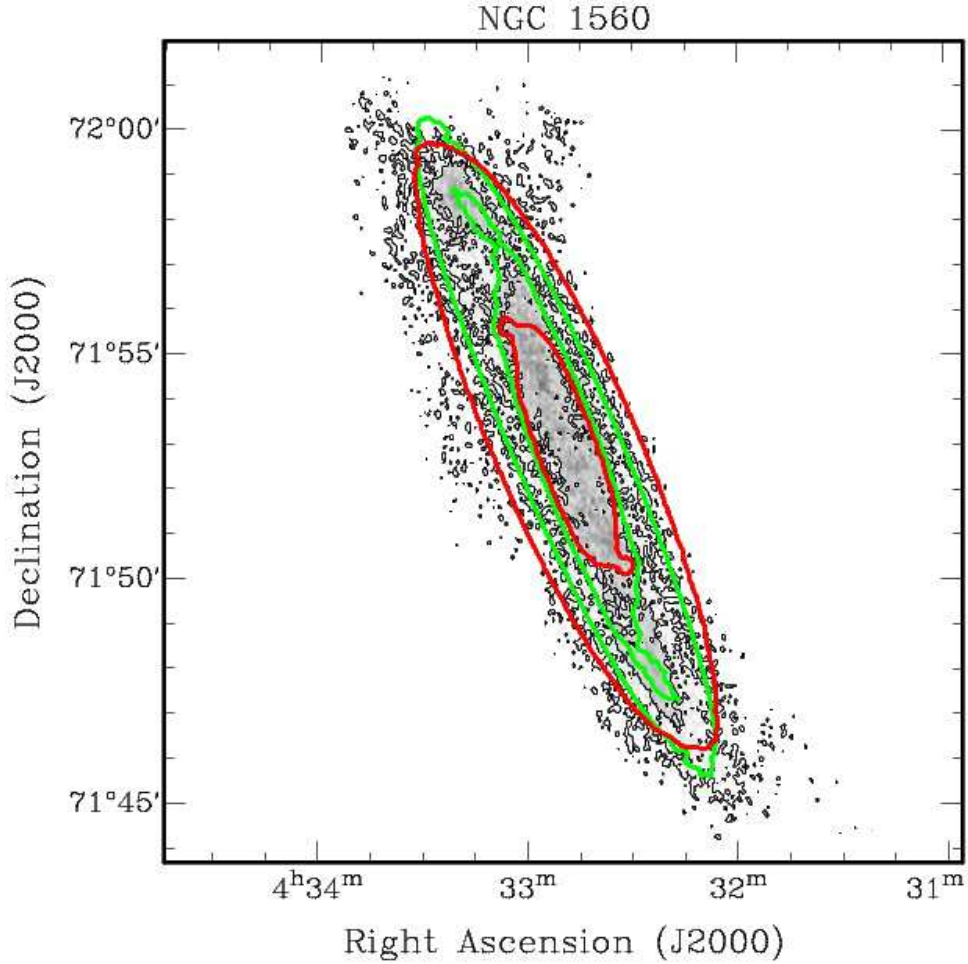}
\caption{Comparison of the observed total HI map (black contours
and greyscale) with the total HI map derived from a model datacube
assuming an inclination angle of 78$^{\circ}$ (red contours)
and 82$^{\circ}$ (green contours). An inclination angle of 82$^{\circ}$
gives a better representation of the observations. Contours
are 1 and 3 $\times 10^{21}$ atom cm$^{-2}$.}
\label{mom0_incl}
\end{figure*}

\begin{figure*}
\includegraphics[scale=1.15]{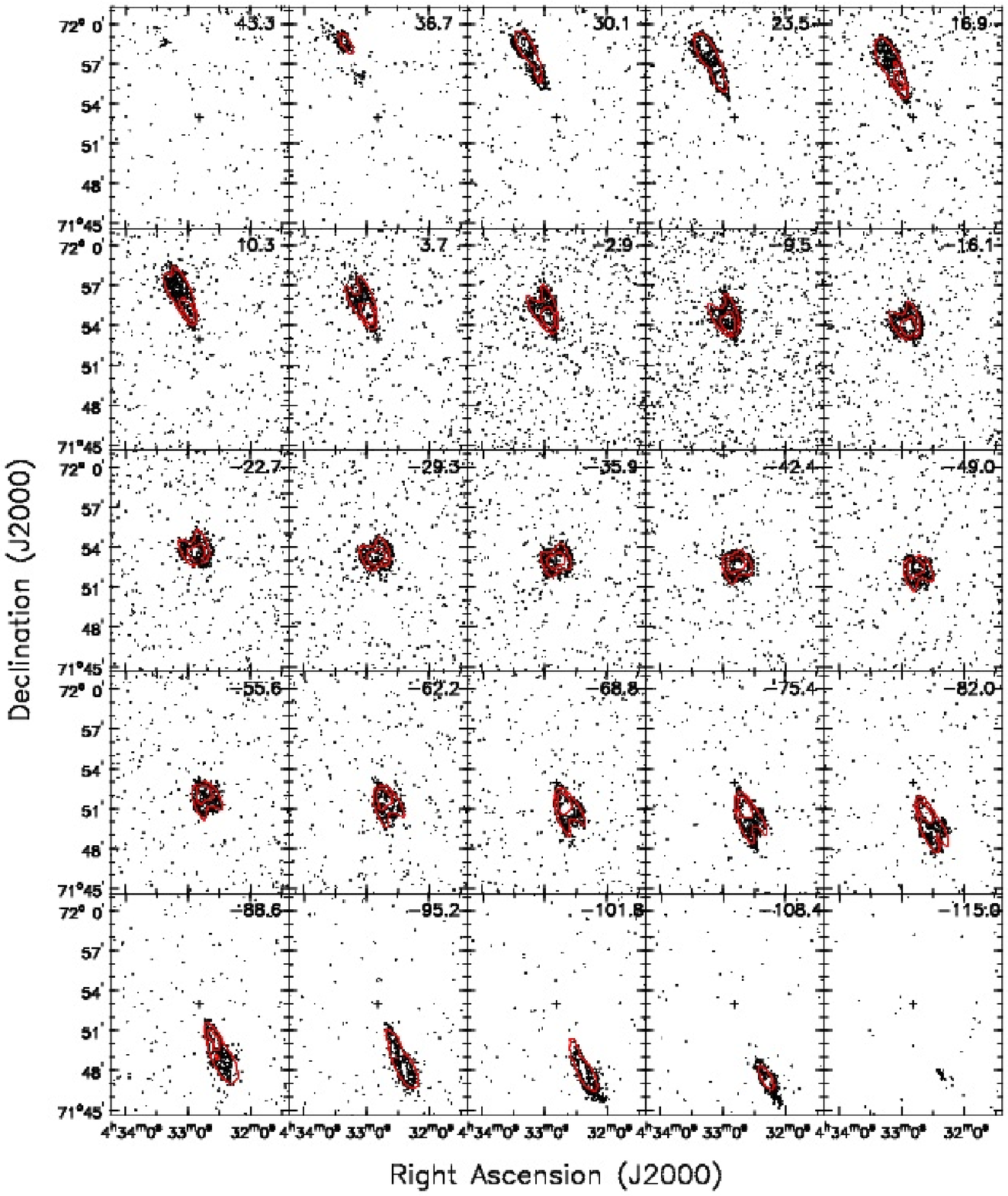}
\caption{Same as Fig. \ref{channels_hires}, with the addition
of red contours, which represent our model datacube.
} 
\label{channels_datiemod.hires}
\end{figure*}

Then, based on the rotation curve, the geometrical parameters 
derived from the tilted-ring modelling of the velocity field, 
on an assumed scale-height of the HI layer of 0.2 kpc (Barbieri 
et al., 2005), an HI
velocity dispersion of 10 km s$^{-1}$ (Tamburro 
et al., 2009), and
the surface density profile, we built model datacubes to check
the validity of our derived parameters. Comparison with the 
data was made channel map by channel map and on the moment maps. 
In particular, it turns out that, in order to reproduce the total
HI map, the inclination angle had to be changed from 78$^\circ$
to 82$^\circ$ (and the rotation velocity was corrected by
a factor sin(78$^\circ$)/sin(82$^\circ$)). This is illustrated
in Fig. \ref{mom0_incl}.
Once this change was made, the agreement between the model
datacube and the observed one is excellent, as can be seen 
in Fig. \ref{channels_datiemod.hires}. Also, contrary to Gentile et al.
(2007a), the central channel maps are well reproduced without the need of
introducing non-circular motions in the model datacube.

\begin{figure}
\includegraphics[scale=0.41]{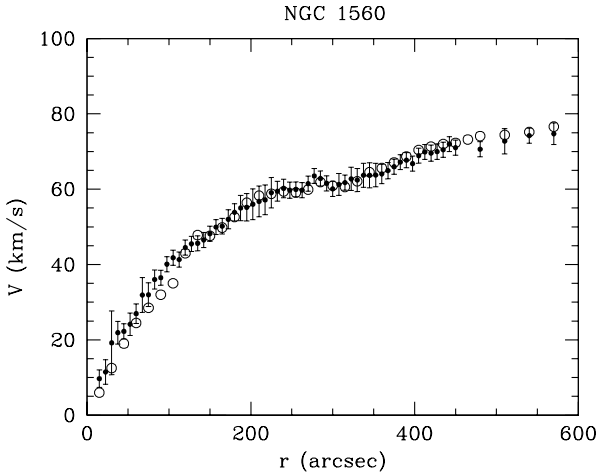}
\caption{Comparison between the rotation curve of Broeils (1992) (empty
circles) and the rotation curve derived in the present paper (full circles).
} 
\label{rcfit.comparisonbr}
\end{figure}

\begin{figure}
\includegraphics[scale=0.43]{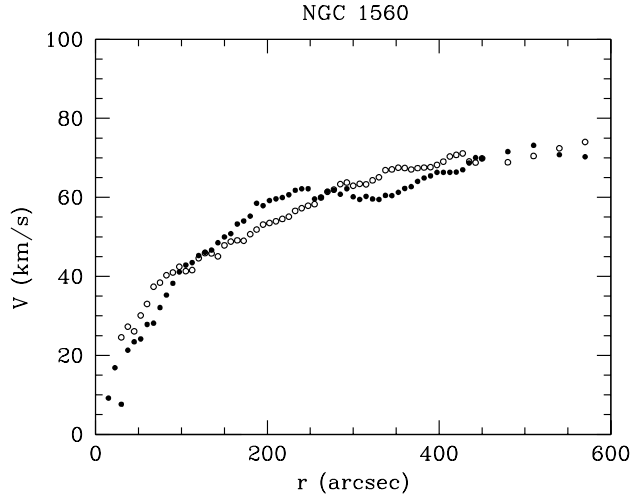}
\caption{
Rotation curve of NGC 1560 determined for the two sides separately. 
Empty circles represent the southern (approaching) side, whereas full
circles represent the northern (receding) side.
} 
\label{rcfit.apprecsep}
\end{figure}

The rotation curve (Fig. \ref{rcfit.comparisonbr}) is very
similar to the one derived by B92. The last four points of the 
rotation curve were derived using the velocity field made from
the low-resolution data cube. The largest differences between our
rotation curve and the one derived by B92 are of order $\lesssim 5$ km s$^{-1}$ 
in the innermost $\sim 100 \arcsec$, where projection effects
are expected to be stronger. Then, before making the mass models,
the rotation curve was corrected for asymmetric drift following
B92 and Skillman et al. (1987). The corrections were smaller than
the errorbars. For the last three data points of the rotation curve
(when the surface density, of at least one side of the galaxy, drops
below 1 M$_{\odot}$ pc$^{-2}$), we did not apply the correction, as
it would be too uncertain because it would imply dividing by
values of the surface density very close to zero.
The errorbars on the rotation curve were calculated from the
difference between the approaching and the receding side; a minimum
realistic error of 2 km s$^{-1}$ was taken into account. The rotation curve
determined separately for the two sides of the galaxy is 
shown in Fig. \ref{rcfit.apprecsep}. Similarly to what B92 had
found, we find that globally the rotation curve is quite symmetric, but that
in the region between 200$\arcsec$ and 350$\arcsec$ the asymmetries are largest, typically 
of order $6-7$ km s$^{-1}$. We note that the wiggle is clearly present on 
one side (the northern side) and barely visible on the other side; in this 
respect, the kinematic asymmetry is very similar to the surface density asymmetry.

\section{Mass models}

\subsection{The contribution of visible matter}
\label{sect_vis}

The contribution of the gaseous disk to the rotation curve
($V_{\rm gas}$) was derived using the task ROTMOD in GIPSY, which makes use
of the method outlined in Casertano (1983). We used
the surface density profile derived in Section \ref{sect_hi}
and we assumed the 
same scale-height as in our model datacubes, i.e. 0.2 kpc.
Different (but realistic) values of the scale-height do not
affect $V_{\rm gas}$ significantly. The HI surface density distribution
was then multiplied by a factor 1.33 to account for primordial helium.

In order to derive the shape of the contribution of the stellar disk 
to the rotation curve
($V_{\rm stars}$), we applied the ROTMOD to the I-band photometric
data obtained by Buta \& McCall (1999). Also in this case, we assumed
a scale-height of 0.2 kpc. 
Using the range of major-axis scale lengths given in Buta \& McCall,
and assuming that the scale length/scale height ratio is 7.3 (Kregel, van der Kruit
\& de Grijs 2002), we find a possible range of scale heights of $0.13 - 0.37$ kpc.  
Again, assuming an infinitely thin
disk or a thicker - but realistic - disk would not significantly change the resulting $V_{\rm stars}$.
The absolute scaling of $V_{\rm stars}$ depends
on the stellar mass-to-light ($M/L$) ratio. One way of estimating its
value is from stellar population synthesis models, which find a correlation
between observed colour and $M/L$ ratio. We used the method described
in Bell \& de Jong (2001), and from the $(V-I)$ colour given in Buta \& 
McCall (1999) we found and I-band ($M/L_{\rm I}$) mass-to-light ratio of 1.43.
A secure assessment of the uncertainty on this value is virtually impossible
to give, because it combines observational and theoretical uncertainties.
We estimate it to be around 0.2 dex (de Jong \& Bell 2007), therefore
in our fits we leave $M/L_{\rm I}$ as a free parameter, constrained within
0.2 dex around 1.43.

\subsection{Dark matter and MOND}

To explain the mass discrepancy in NGC 1560, one has to resort to
either a dark matter halo or MOND.

For the dark matter halo, we considered two different possibilities:
a Burkert halo and an NFW halo (see also Section \ref{sect_results} for a discussion of the Einasto halo). The Burkert halo (Burkert 1995, Salucci
\& Burkert 2001) is an empirical functional form for the density
distribution of dark matter in galaxies ($\rho_{\rm Bur}$), 
which generally gives
good fits to rotation curves:

\begin{equation}   
\rho_{\rm Bur}(r)=\frac{\rho_0 r_{\rm core}^3}{(r+r_{\rm core})   
(r^2+r_{\rm core}^2)}   
\end{equation}   
   
\noindent    
were $\rho_0$ is the central density and $r_{\rm core}$ is the core 
radius. The Burkert halo has a constant density core at the centre.

The NFW halo (Navarro, Frenk \& White 1996) is the result of an 
analytical fit to the dark matter density distribution that comes out
of cosmological simulations performed within the frame of the ($\Lambda$)CDM
theory. The density distribution $\rho_{\rm NFW}(r)$ is given by 

\begin{equation}   
\rho_{\rm NFW}(r)=\frac{\rho_{\rm s}}{(r/r_{\rm s})(1+r/r_{\rm s})^2},    
\end{equation}   
   
\noindent    
where $\rho_{\rm s}$ and $r_{\rm s}$ are the characteristic density and    
scale of the NFW halo. A more useful pair of parameters can be found
in the concentration parameter ($c_{\rm vir}$) and 
the virial mass ($M_{\rm vir}$). Cosmological simulations show that these 
two parameters are in fact correlated (Bullock et al. 2001,
Wechsler et al. 2002, Neto et al. 2007), so that the following relations
apply: 

\begin{equation}   
c_{\rm vir} \simeq 13.6 \left( \frac{M_{\rm vir}}{10^{11}{\rm M}_{\odot}} \right)^{-0.13}     
\label{cmvir}   
\end{equation}   
 
\begin{equation}  
r_{\rm s} \simeq 8.8 \left( \frac{M_{\rm vir}}{10^{11}{\rm M}_{\odot}}   
\right)^{0.46} {\rm kpc}   
\end{equation}   
   
\begin{equation}   
\rho_{\rm s} \simeq \frac{\Delta}{3}    
\frac{c_{\rm vir}^3}{{\rm ln}(1+c_{\rm vir})-c_{\rm vir}/(1+c_{\rm vir})} \rho_{\rm crit}      
\label{cmvir2}   
\end{equation}   
   
\noindent    

where $\Delta$ is the virial overdensity at $z=0$; it can be calculated
following Bryan \& Norman (1998).

An alternative explanation to the presence of dark matter in galaxies
is MOND. In MOND, the true gravitational acceleration $\vec{g}$ can
be computed from the Newtonian acceleration $\vec{g_{\rm N}}$ through
the following relation:

\begin{equation}
\vec{g} = \vec{g_{\rm N}}/\mu(|g|/a_0)
\end{equation}

where $a_0 \sim 1.2 \times 10^{-8}$ cm s$^{-2}$ (Begeman, Broeils \& Sanders 
1991), and $\mu(x)$ is the so-called interpolating function, whose
asymptotic values are $\mu(x) = x$ when $x \ll 1$ and $\mu(x) = 1$ when
$x \gg 1$. The exact functional form of $\mu(x)$ is not defined by MOND,
and we adopt here the ``simple'' interpolating function (Famaey \& Binney 2005,
Zhao \& Famaey 2006):

\begin{equation}
\mu(x) = \frac{x}{1+x}  
\label{mond_simple}   
\end{equation}

which has been shown to yield more realistic fits than the ``standard'' 
$\mu(x)$ (Sanders \& Noordermeer 2007, Famaey et al. 2007b, 
Angus, Famaey \& Diaferio 2010), which has the following form:

\begin{equation}   
\mu_{\rm standard}(x)=\frac{x}{\sqrt{1+x^2}}   
\label{mond_std}   
\end{equation}

Because the estimates of the distance of NGC 1560 span a large range
of values in the literature (see Section 1), in the MOND fits we decided
to leave it as a free parameter, checking {\it a posteriori} the validity
of the best-fit value.

\section{Mass modelling results}
\label{sect_results}

\begin{figure}
\includegraphics[scale=0.41]{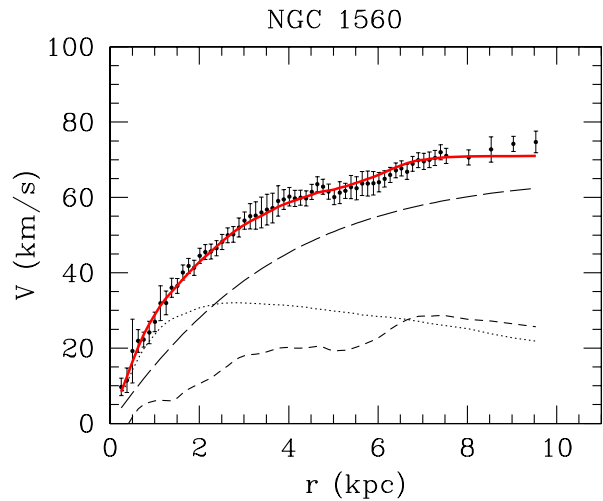}
\caption{Rotation curve fit using the Burkert halo. 
Short-dashed, dotted, and long-dashed lines
represent the Newtonian contributions of the gaseous disk, stellar disk,
and dark halo, respectively. The best-fit model is shown as a solid red line.
} 
\label{rcfit.bur}
\end{figure}

\begin{figure}
\begin{center}
\includegraphics[scale=0.41]{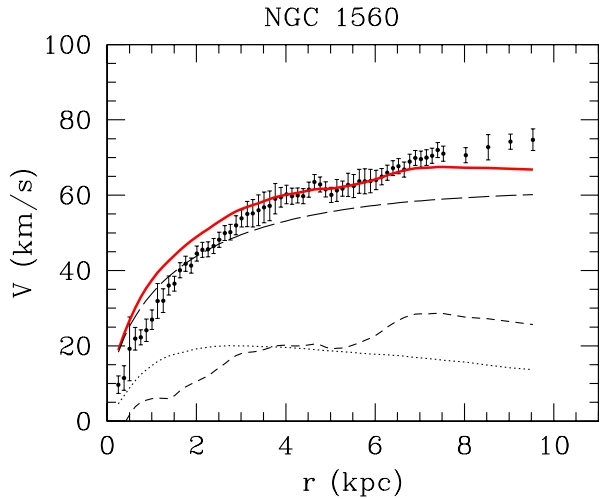}
\end{center}
\caption{Rotation curve fit using the NFW halo. 
Lines and symbols are like those in Fig. \ref{rcfit.bur}.
} \label{rcfit.nfw}
\end{figure}

\begin{figure}
\begin{center}
\includegraphics[scale=0.41]{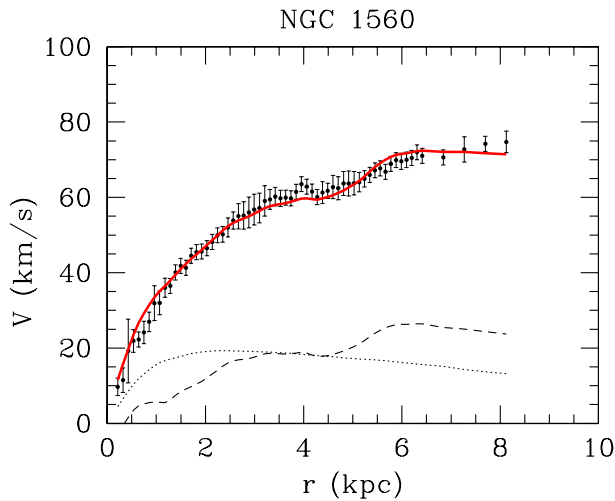}
\end{center}
\caption{Rotation curve fit using MOND. The best-fit distance is 2.94 Mpc.
Lines and symbols are like those in Fig. \ref{rcfit.bur}.
} \label{rcfit.1560.simple}
\end{figure}

Figures \ref{rcfit.bur}-\ref{rcfit.1560.simple} show the mass modelling results.
The Burkert halo gives a very good fit to the rotation curve ($\chi^2_{\rm red}=0.33$),
with a core radius of 5.6 kpc and a central density of 0.8$\times$10$^{-24}$ g cm$^{-3}$.  
However, because of the halo dominance already at small radii, it does not manage to explain
the ``wiggle'' around 300$\arcsec$: the total rotation curve that
results from the mass modelling with a Burkert halo is featureless, whereas
the observed rotation curve is not. Note, however, that the best-fit curve
goes through the (conservative) errorbars in the region of the wiggle. 
The best-fit stellar $M/L_{\rm I}$ ratio is
2.3, at the high end of the allowed range (see Section \ref{sect_vis}).

On the other hand, modelling the rotation curve using the halo predicted
in $\Lambda$CDM simulations results in a bad quality fit 
(Fig. \ref{rcfit.nfw}). The best-fit virial mass is ($4.4\pm0.4$)$\times$10$^{10}$ M$_\odot$
(the concentration is derived through eq. \ref{cmvir}),
which is consistent with studies linking the stellar and dark halo
masses (e.g. Shankar et al. 2006, Guo et al. 2010).
As in numerous galaxy rotation curves,
the observed shape of the rotation curve is very different from the 
one predicted using an NFW halo, in particular in the innermost parts.
The best-fit stellar $M/L_{\rm I}$ ratio is 0.9, which is at the lower extreme 
of the range of $M/L_{\rm I}$ we considered.
The quality of the NFW can improve if we take both the concentration
and the virial mass as free parameters. However, the price to pay is to 
have a best-fit virial mass of 3.0 $\times$ 10$^{11}$ M$_{\odot}$,
which is much too high for a galaxy with a stellar mass around 5$\times 10^{8}$
M$_{\odot}$ (Shankar et al. 2006, Guo et al. 2010), and 
to have a best-fit concentration parameter
of 6.1, which is 2 $\sigma$ to 2.5 $\sigma$ below the scatter in the 
virial mass-concentration relation (eq. \ref{cmvir})
found in $\Lambda$CDM simulations (Bullock et al. 2001, Neto et al. 2007).

The so-called Einasto halo (Einasto 1965, Navarro et al. 2004, 
Navarro et al. 2009), which is a functional form that gives a slightly
better description of the density distribution simulated halos than the NFW 
formula, was not used here. The reason is that within the radial range
probed by our data (0.02$r_{\rm max}$ to 0.73$r_{\rm max}$, where
$r_{\rm max}$=2.16$r_{\rm s}$ and $r_{\rm s}$ is derived from the 
best-fit $M_{\rm vir}$ (with the concentration fixed) and 
eqs. \ref{cmvir}-\ref{cmvir2}), following
Navarro et al. (2004, 2009) we find that the velocity difference between the 
two profiles is $\lesssim 0.1$dex, and it would make the velocity in the 
innermost parts higher, so the agreement with the data would be even {\it worse}.

MOND fits the rotation curve very well; we recall that we use the
simple interpolating function (eq. \ref{mond_simple}). 
Formally, the reduced $\chi^2$ value (0.56)
is a little higher than the Burkert halo fits, but because the ``wiggle'' 
around 300$\arcsec$ appears both in the total rotation curve and in 
$V_{\rm gas}$, the MOND fit reproduces reasonably well the wiggle.  
The best-fit stellar $M/L_{\rm I}=0.98$ lies within the range derived from
stellar population synthesis. The best-fit distance (2.94 Mpc) is 
a little bit low compared with the value given in Karachentsev et al. (2003),
i.e. 3.45 Mpc. However, the quality of the fit only slightly decreases
(reduced $\chi^2$=0.69) if we force the distance to stay within
the range allowed by Karachentsev et al. (2003). Also, we note that there
are also lower estimates of NGC 1560's distance, e.g. Lee \& Madore (1993),
who give 2.5 Mpc using the brightest stars method. 
Using the standard interpolating function (eq. \ref{mond_std}) gives a slightly
higher best-fit distance (3.16 Mpc) and stellar $M/L_{\rm I}$ ratio (1.09),
for an equivalently good fit ($\chi^2$=0.54, not shown here).

It has been noted in the past (e.g. Bosma 1999) that there are cases where wiggles are linked to non-circular motions due to spiral arms. Obviously, this effect is much more prominent in long-slit data than in two-dimensional velocity fields. Bosma mentions Visser (1980), where in M81 strong non-circular motions in a 2-D velocity field still have a (small) effect on the rotation curve. However, in the case of NGC 1560 there are non-prominent spiral arms, so we expect the effect on the rotation curve to be very small, certainly not as large as the observed wiggle.

\begin{figure*}
\begin{center}
\includegraphics[scale=0.82]{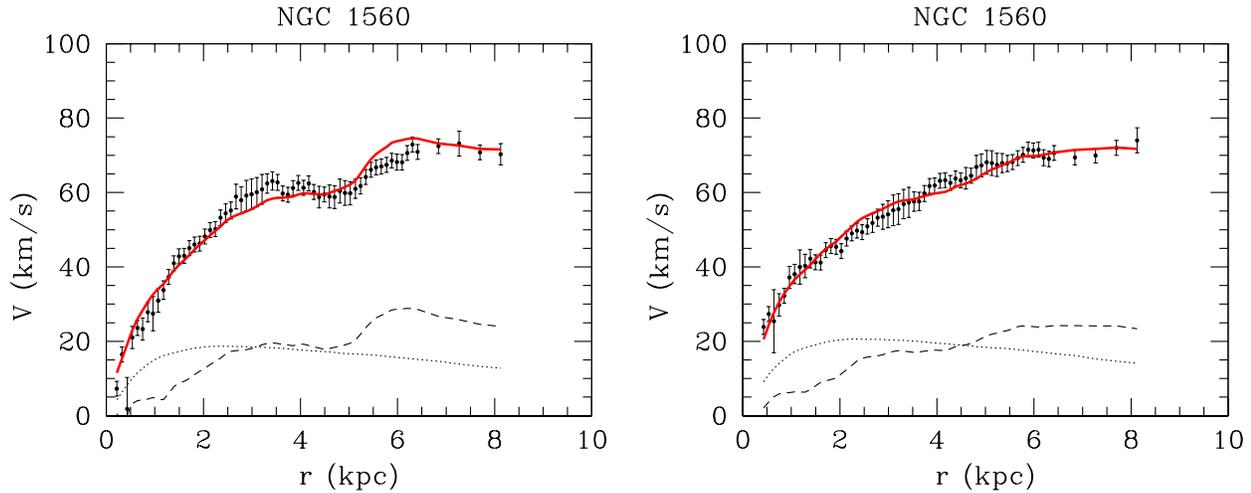}
\end{center}
\caption{Rotation curve fits using MOND, fitting separately the two
sides of the galaxy and using the simple interpolation function. 
On the left is the northern (receding) side of the galaxy,
on the right is the southern (approaching) side. 
Lines and symbols are like those in Fig. \ref{rcfit.bur}.
} \label{combined_separately_standard}
\end{figure*}

At the radius of the wiggle, the orbital frequency is $\sim$14 km s$^{-1}$ 
kpc$^{-1}$ and the epicyclic frequency is $\sim$21 km s$^{-1}$ kpc$^{-1}$, which 
means that the mass distribution has the time to react to the gravitational 
potential from one side of the galaxy to the other, but only barely. Hence, it
is interesting (though not formally completely correct, the construction of
a rigorous model goes beyond the scope of this paper) to make mass models of the two sides 
(approaching and receding) of the galaxy independently, as if the two sides
were separate and in independent circular motion. We kept the distance fixed
at 2.94 Mpc, the best-fit distance in the total MOND fit.  The results are shown in
Fig. \ref{combined_separately_standard}, where one can notice that the observed
kinematics follows the distribution of baryons, even when the two sides are considered
separately: in the receding (northern) side of the galaxy, the wiggle in the baryons
distribution is much more pronounced, and this is reflected in the observed kinematics
of that side of the galaxy.

%
   
\section{Conclusions}

NGC 1560 is a nearby dwarf Sd galaxy, whose rotation curve has 
a very distinct ``wiggle''. We observed NGC 1560 in HI with
the GMRT, achieving a two times better resolution than the 
previous data of Broeils (1992), which were obtained with the WSRT.

We re-derived the rotation curve of NGC 1560 by taking projection
effects into account (because of its high inclination angle, 
$\sim 80^\circ$), and checked the reliability of our findings by creating
model datacubes, which were compared to the observations.

The new rotation curves is similar to the one derived by Broeils (1992),
the main differences being in the innermost $\sim 100 \arcsec$: at those
radii we find slightly ($\lesssim 5$ km s$^{-1}$) higher velocities than
Broeils (1992). Also, we confirm the presence of a ``wiggle'' in the rotation
curve, at around 300$\arcsec$.

The rotation curve was then corrected for asymmetric drift and 
used as input for mass modelling. The contribution of the stellar
disk to the rotation curve was derived from NIR (I-band) data.
The core-dominated Burkert halo gives a good fit to the observed
rotation curve, but it does not manage to explain the wiggle.
The NFW halo gives a bad fit, greatly overpredicting the velocity in 
the innermost regions and slightly underpredicting the outermost
ones; using an Einasto halo would only slightly change the fits,
making them marginally even worse. MOND gives a very good account of 
the data, particularly of the wiggle. 

HI observations at about twice the spatial resolution of the previous
ones confirmed thus that NGC 1560 is a nice example of the connection
between baryons and total kinematics in galaxies (an expression of 
which is MOND).

\section*{acknowledgements}

GG and KVA are postdoctoral researchers of the FWO-Vlaanderen (Belgium).
BF is a Senior Research Associate of the CNRS (France). 
We thank the referee, Stacy McGaugh, for insightful comments that
improved the quality of this paper.
We thank the staff of the GMRT who have made these observations possible. GMRT is run by the National Centre for Radio Astrophysics  of the Tata Institute of Fundamental Research.

\label{lastpage}


\begin{thebibliography}{}

\bibitem[Angus et al.(2007)]{2007ApJ...654L..13A} Angus, G.~W., Shan, H.~Y., Zhao, H.~S., \& Famaey, B.\ 2007, ApJ, 654, L13 
\bibitem[Angus et al.(2010)]{2010MNRAS.402..395A} Angus, G.~W., Famaey, B., 
\& Diaferio, A.\ 2010, MNRAS, 402, 395 
\bibitem[Barbieri et 
al.(2005)]{2005A&A...439..947B} Barbieri, C. V., Fraternali, F., Oosterloo, T., Bertin, G., Boomsma, R., \& Sancisi, R.\ 2005, A\&A, 439, 947 
\bibitem[\protect\citeauthoryear{Begeman}{1989}]{B:89}   
 Begeman, K. G., 1989, A\&A, 223, 47  
\bibitem[\protect\citeauthoryear{Begeman, Broeils \& Sanders}{Begeman et al.}{19   
91}]{Beg:91} Begeman, K. G., Broeils, A. H., Sanders, R. H., 1991, MNRAS, 249, 5   
23  
\bibitem[\protect\citeauthoryear{Bell \& de Jong}{2001}]{BdJ:01} Bell, E. F., de   
 Jong, R. S., 2001, ApJ, 550, 212  
 \bibitem[Bienaym{\'e} et al.(2009)]{2009A&A...500..801B} Bienaym{\'e}, O., Famaey, B., Wu, X., Zhao, H.~S., \& Aubert, D.\ 2009, A\&A, 500, 801 
\bibitem[Blumenthal et al.(1986)]{1986ApJ...301...27B} Blumenthal, G.~R., Faber, S.~M., Flores, R., \& Primack, J.~R.\ 1986, ApJ, 301, 27 
\bibitem[Bosma(1999)]{1999ASPC..182..339B} Bosma, A.\ 1999, Galaxy Dynamics 
- A Rutgers Symposium, 182, 339 
\bibitem[Broeils(1992)]{1992A&A...256...19B} Broeils, A.~H.\ 1992, A\&A, 256, 19 
\bibitem[\protect\citeauthoryear{Bryan \& Norman}{1998}]{Bry:98} Bryan, G. L., Norman,   
M. L., 1998, ApJ, 495, 80 
\bibitem[\protect\citeauthoryear{Bullock et al.}{2001}]{Bul:01} Bullock, J.S., Kolatt, T.S., Rachel, Y.S., Somerville, S., Kravtsov, A.V., Klypin, A.A., Primack, J.R., Dekel, A., 2001, MNRAS, 321, 559 
\bibitem[\protect\citeauthoryear{Burkert}{1995}]{B:95} Burkert, A., 1995, ApJ,   
  447, L25  
\bibitem[Corbelli et al.(2009)]{2009arXiv0912.4133C} Corbelli, E., Lorenzoni, S., Walterbos, R.~A.~M., Braun, R., \& Thilker, D.~A.\ 2010, A\&A, in press (arXiv:0912.4133)
\bibitem[\protect\citeauthoryear{de Blok, McGaugh \& Rubin}{2001}]{dB:01} de Blok, W. J. G., McGaugh, S. S., Rubin, V. C., 2001, AJ, 122, 2396    
\bibitem[de Jong \& Bell(2007)]{2007iuse.book..107D} de Jong, R.~S., \& Bell, E.~F.\ 2007, Island Universes - Structure and Evolution of Disk Galaxies, 107 
\bibitem[de Vaucouleurs(1991)]{1991Sci...254.1667D} de Vaucouleurs, G., et al. 1991, Revised Catalogue of Galaxies Version 3.9 (RC3.9)
\bibitem[Donato et al.(2009)]{2009MNRAS.397.1169D} Donato, F., et al.\ 2009, MNRAS, 397, 1169 
\bibitem[Einasto (1965)]{Einasto1965} Einasto J., 1965, Trudy Inst. Astroz. Alma-Ata, 51, 87
\bibitem[Famaey \& Binney(2005)]{2005MNRAS.363..603F} Famaey, B., \& Binney, J.\ 2005, MNRAS, 363, 603 
\bibitem[Famaey et al.(2007)]{2007MNRAS.377L..79F} Famaey, B., Bruneton, J.-P., \& Zhao, H.\ 2007a, MNRAS, 377, L79 
\bibitem[Famaey et al.(2007)]{2007PhRvD..75f3002F} Famaey, B., Gentile, G., Bruneton, J.-P., \& Zhao, H.\ 2007b, Phys. Rev. D, 75, 063002 
\bibitem[\protect\citeauthoryear{Gentile et al.}{2004}]{Ge:04} Gentile, G., Salucci, P., Klein, U., Vergani, D., Kalberla, P., 2004, MNRAS, 351, 903   
\bibitem[\protect\citeauthoryear{Gentile et al.}{2005}]{Ge:05}   
Gentile, G., Burkert, A., Salucci, P., Klein, U., Walter, F., 2005, ApJ, 634,   
L145   
\bibitem[Gentile et al.(2007)]{2007MNRAS.375..199G} Gentile, G., Salucci, P., Klein, U., \& Granato, G.~L.\ 
2007a, MNRAS, 375, 199
 \bibitem[Gentile et al.(2007)]{2007A&A...472L..25G} Gentile, G., Famaey, B., Combes, F., Kroupa, P., Zhao, H.~S., \& Tiret, O.\ 2007b, A\&A, 472, L25 
\bibitem[Gentile(2008)]{2008ApJ...684.1018G} Gentile, G.\ 2008, ApJ, 684, 1018 
\bibitem[Gentile et al.(2009)]{2009Natur.461..627G} Gentile, G., Famaey, B., Zhao, H., \& Salucci, P.\ 2009, Nature, 461, 627 
\bibitem[Gnedin et al.(2004)]{2004ApJ...616...16G} Gnedin, O.~Y., Kravtsov, A.~V., Klypin, A.~A., \& Nagai, D.\ 2004, ApJ, 616, 16
\bibitem[Governato et al.(2009)]{2009arXiv0911.2237G} Governato, F., et al.\ 2010, Nature, in press (arXiv:0911.2237) 
\bibitem[Greisen et al.(2009)]{2009AJ....137.4718G} Greisen, E.~W., Spekkens, K., \& van Moorsel, G.~A.\ 2009, AJ, 137, 4718 
\bibitem[Guo et al.(2010)]{2010MNRAS.tmp..367G} Guo, Q., White, S., Li, C., 
\& Boylan-Kolchin, M.\ 2010, MNRAS, 367 
\bibitem[Karachentsev et 
al.(2003)]{2003A&A...408..111K} Karachentsev, I.~D., Sharina, M.~E., Dolphin, A.~E., \& Grebel, E.~K.\ 2003, A\&A, 408, 111 
\bibitem[Kregel et al.(2002)]{2002MNRAS.334..646K} Kregel, M., van der Kruit, P.~C., \& de Grijs, R.\ 2002, MNRAS, 334, 646 
\bibitem[Krismer et al.(1995)]{1995AJ....110.1584K} Krismer, M., Tully, R.~B., \& Gioia, I.~M.\ 1995, AJ, 110, 1584 
\bibitem[Kuzio de Naray et al.(2006)]{2006ApJS..165..461K} Kuzio de Naray, R., McGaugh, S.~S., de Blok, W.~J.~G., \& Bosma, A.\ 2006, ApJS, 165, 461 
\bibitem[Lee \& Madore(1993)]{1993AJ....106...66L} Lee, M.~G., \& Madore, B.~F.\ 1993, AJ, 106, 66 
\bibitem[Marchesini et al.(2002)]{2002ApJ...575..801M} Marchesini, D., D'Onghia, E., Chincarini, G., Firmani, C., Conconi, P., Molinari, E., \& Zacchei, A.\ 2002, ApJ, 575, 801 
\bibitem[Mashchenko et al.(2006)]{2006Natur.442..539M} Mashchenko, S., Couchman, H.~M.~P., \& Wadsley, J.\ 2006, Nature, 442, 539 
\bibitem[McGaugh \& de Blok(1998)]{1998ApJ...499...66M} McGaugh, S.~S., \& de Blok, W.~J.~G.\ 1998, ApJ, 499, 66
\bibitem[McGaugh et al.(2000)]{2000ApJ...533L..99M} McGaugh, S.~S., Schombert, J.~M., Bothun, G.~D., \& de Blok, W.~J.~G.\ 2000, ApJ, 533, L99 
\bibitem[McGaugh(2005)]{2005PhRvL..95q1302M} McGaugh, S.~S.\ 2005a, Phys. Rev. Letters, 95, 171302 
\bibitem[McGaugh(2005)]{2005ApJ...632..859M} McGaugh, S.~S.\ 2005b, ApJ, 632, 859 
\bibitem[McMillan \& Dehnen(2005)]{2005MNRAS.363.1205M} McMillan, P.~J., \& Dehnen, W.\ 2005, MNRAS, 363, 1205 
\bibitem[\protect\citeauthoryear{Milgrom}{1983}]{Mi:83} Milgrom, M., 1983, ApJ, 270, 365   
\bibitem[Milgrom \& Sanders(2003)]{2003ApJ...599L..25M} Milgrom, M., \& Sanders, R.~H.\ 2003, ApJ, 599, L25 
\bibitem[\protect\citeauthoryear{Navarro et al.}{1996}]{NFW:96} Navarro, J.F., Frenk, C.S., White, S.D.M., 1996, ApJ, 462, 563 
\bibitem[Navarro et al.(2004)]{2004MNRAS.349.1039N} Navarro, J.~F., et al.\ 2004, MNRAS, 349, 1039 
\bibitem[Navarro et al.(2009)]{2009MNRAS.tmp.1918N} Navarro, J.~F., et al.\ 2009, MNRAS, 1918 
\bibitem[Neto et al.(2007)]{2007MNRAS.381.1450N} Neto, A.~F., et al.\ 2007, MNRAS, 381, 1450    
\bibitem[Persic et al.(1996)]{1996MNRAS.281...27P} Persic, M., Salucci, P., \& Stel, F.\ 1996, MNRAS, 281, 27
\bibitem[Salucci \& Burkert(2000)]{2000ApJ...537L...9S} Salucci, P., \& Burkert, A.\ 2000, ApJ, 537, L9 
\bibitem[Salucci et al.(2007)]{2007MNRAS.378...41S} Salucci, P., Lapi, A., Tonini, C., Gentile, G., Yegorova, I., \& Klein, U.\ 2007, MNRAS, 378, 41 
\bibitem[Sancisi \& Allen(1979)]{1979A&A....74...73S} Sancisi, R., \& Allen, R.~J.\ 1979, A\&A, 74, 73 
\bibitem[Sancisi(2004)]{2004IAUS..220..233S} Sancisi, R.\ 2004, Dark Matter in Galaxies, 220, 233 
\bibitem[Sandage(1988)]{1988ApJ...331..605S} Sandage, A.\ 1988, ApJ, 331, 605 
\bibitem[Sanders \& Noordermeer(2007)]{2007MNRAS.379..702S} Sanders, R.~H., \& Noordermeer, E.\ 2007, MNRAS, 379, 702 
\bibitem[Sellwood \& McGaugh(2005)]{2005ApJ...634...70S} Sellwood, J.~A., \& McGaugh, S.~S.\ 2005, ApJ, 634, 70 
\bibitem[Sellwood(2008)]{2008ApJ...679..379S} Sellwood, J.~A.\ 2008, ApJ, 679, 379 
\bibitem[Shankar et al.(2006)]{2006ApJ...643...14S} Shankar, F., Lapi, A., 
Salucci, P., De Zotti, G., \& Danese, L.\ 2006, ApJ, 643, 14 
\bibitem[Skillman et al.(1987)]{1987A&A...185...61S} Skillman, E.~D., Bothun, G.~D., Murray, M.~A., \& Warmels, R.~H.\ 1987, A\&A, 185, 61 
\bibitem[Tamburro et al.(2009)]{2009AJ....137.4424T} Tamburro, D., Rix, 
H.-W., Leroy, A.~K., Low, M.-M.~M., Walter, F., Kennicutt, R.~C., Brinks, 
E., \& de Blok, W.~J.~G.\ 2009, AJ, 137, 4424 
\bibitem[Tiret et al.(2007)]{2007A&A...476L...1T} Tiret, O., Combes, F., Angus, G.~W., Famaey, B., \& Zhao, H.~S.\ 2007, A\&A, 476, L1 
\bibitem[Verheijen \& Sancisi(2001)]{2001A&A...370..765V} Verheijen, M.~A.~W., \& Sancisi, R.\ 2001, A\&A, 370, 765 
\bibitem[Visser(1980)]{1980A&A....88..149V} Visser, H.~C.~D.\ 1980, A\&A, 88, 149 
\bibitem[\protect\citeauthoryear{Wechsler et al.}{2002}]{W:02} Wechsler, R. H., Bullock, J. S., Primack, J. L., Kravtsov, A. V., Dekel, A., 2002, ApJ, 568, 52   
\bibitem[Weinberg \& Katz(2002)]{2002ApJ...580..627W} Weinberg, M.~D., \& Katz, N.\ 2002, ApJ, 580, 627 
\bibitem[Zhao \& Famaey(2006)]{2006ApJ...638L...9Z} Zhao, H.~S., \& Famaey, B.\ 2006, ApJ, 638, L9 


\end{thebibliography}
\end{document}